\documentclass[
reprint, raggedbottom, amsmath,amssymb, aps, pre, nofootinbib, floatfix,
]{revtex4-2}

\usepackage{booktabs}
\usepackage{orcidlink}
\usepackage{graphicx}
\usepackage{dcolumn}
\usepackage{bm}
\usepackage{tikz}
\usepackage{hyperref}
\usepackage[english]{babel}

\graphicspath{{Figures/}}

\begin{document}

\title{Spatial scaling of network length and shared-node density in bilayer distribution grids}

\author{Emile Emery\,\orcidlink{0000-0001-5074-1163}}
\author{Jos\'e Halloy\,\orcidlink{0000-0003-1555-2484}}
\affiliation{Universit\'e Paris Cit\'e, CNRS, LIED UMR 8236, Paris, F-75006, France}
\date{\today}

\begin{abstract}
Certain distribution networks are built in layers, each layer
collecting a commodity from the layer above it and delivering it to the
layer below. We propose a minimal model that gives the length of each
layer per unit area from the local density of demand, here the
population density. Starting from the
Beardwood, Halton and Hammersley law for minimal networks, the length of a
layer depends on the number of nodes it reaches and on the area occupied
by the users. When two consecutive layers share this area, the gap between
their two length exponents is fixed by the exponent of the density of
their shared nodes, with no adjusted parameter.
We test the model on the complete low and medium-voltage grid of
metropolitan France, at the resolution of a population raster. The
exponents are fitted by Poisson maximum likelihood on every pixel, with
confidence intervals from a spatial block bootstrap. The length exponents
imply an \emph{effective settled area} that grows sublinearly with the
density, and the gap between them agrees with the exponent of the
substation density, counted independently. The agreement holds over the
whole territory, so that the model can serve to estimate the length of
such networks where only the density of demand is known.
\end{abstract}

\maketitle

\section{Introduction}
\label{sec:intro}
Many natural and engineered systems, from neural systems to roads, pipes and
power lines, are spatial networks, in which nodes and links are located in
physical space~\cite{Barthelemy2011}. This spatial embedding has a strong
effect on the structure of the network. Road networks, for instance, grow by
exploration of new land and densification of settled areas~\cite{Strano2012}.
In this paper, we consider distribution networks, which carry a commodity from
a small number of sources to a large number of users, and for which link
length contributes to construction and maintenance
costs~\cite{GastnerNewmanJSTAT}. Because a longer link costs more to build and
to maintain, the geometry of these networks results from a trade-off between
the cost of their links and the benefit they bring~\cite{LoufBarthelemy2013}.
Arguments of this kind predict the size of a network from the size of what it
serves. They account, for example, for the metabolic rate of
organisms~\cite{WBE1997,WBE1999,WestBrown2004,BMR1999} or for the sublinear
scaling of urban infrastructure with city
size~\cite{Bettencourt2007,Bettencourt2013,LoufRoth2014}.

These laws relate totals, such as the mass of an organism or the population of
a city, and do not say how the density of a network varies from place to place
with the local density of what it serves. Gusein-Zade, and later Gastner and
Newman, showed that the density of facilities that minimizes the mean distance
to the population scales as the $2/3$ power of the population
density~\cite{GuseinZade1982,GastnerNewmanPRE}. Measured values, however,
yield $2/3$ for public services and $1$ for profit-driven facilities, in the
United States as well as in South
Korea~\cite{GuseinZade1993,um_scaling_2009,xu_deconstructing_2020}.

Real distribution networks are sometimes built in levels, each level
collecting the commodity from the one above it and delivering it to the one
below. Networks made of several interacting layers have a general
formalism~\cite{Kivela2014,DeDomenico2013}. In this paper, we use only the
simplest case, two layers that share a set of nodes. In a distribution
network, the layers often come with an order, since each one is served by the
layer above it, and we call such a network multilevel.

We derive a relation between layer lengths and node densities under explicit
geometric assumptions, then ask whether it is compatible with exponents
estimated on an observed grid. The empirical model uses population density as
a demand field, and its amplitudes and exponents require network data for
calibration. Population is a proxy for the locations of users, rather than a
direct measurement of electrical load.

In this paper, we focus on the electricity distribution network, which
comprises medium-voltage (MV, $1$--$50\,$kV) and low-voltage (LV, $<1\,$kV)
lines. It forms the most spatially extensive component of the power system,
branching into every populated area. K\"{u}hnert~\textit{et~al.}\ related the
length of the electrical cables of German cities to their population, not to
the area over which this population is spread~\cite{Kuhnert2006}. Like the
laws above, this relation concerns the totals of whole cities, and it leaves
open how the length varies inside a territory with the local density. The two
levels of this network live at very different scales. The MV layer covers the
whole territory, and an MV/LV substation lies a few kilometers along the lines
from its source substation. The LV layer follows the streets, and the farthest
point of the network of a substation lies a few hundred meters from it. A test
of the theory, therefore, requires a complete map of both layers rather than a
sample. The French distribution network provides such a map, since the LV and
MV lines of nearly all distribution operators and the substations that join
the two layers are recorded over mainland
France~\cite{ore_lines_opendata,ore_substations_opendata}. Complex-network
studies of power grids have mostly worked on a single voltage
level~\cite{PaganiAiello2013}, although the levels differ enough to be read as
the layers of one multilayer network~\cite{Hartmann2021,Emery2024}. In a power
grid, each voltage level is a layer, and the substations are the nodes shared
by two consecutive layers~\cite{nardelli_models_2014, emery_statistical_2025}.

Growth models of transmission grids place the nodes at random or where the
load is highest, and connect each new node to nearby ones with a rule that
favors short links~\cite{Mei2011,Schultz2014,RosasCasals2018,Emery2025}. They
leave aside the distribution layers, where most of the length lies~\cite{IEA},
and where each line serves consumers spread over the territory rather than a
few large substations. The length of a distribution grid is then computed by
engineering tools that plan it, customer by
customer~\cite{Domingo2011,Mateo2020}, or estimated using a model that
predicts substation locations from population density and connects them with a
spanning tree~\cite{Emery2026}. In neither case is the relation between length
and density derived. The present work complements these approaches by
examining the relation between the scaling of two voltage layers and that of
their shared nodes.

Our aim is to create a minimal model that provides the length of each layer
based solely on the demand density field. For the electricity network, we
consider the population density as this field. Population density is used
because gridded population fields are available
globally~\cite{wang_projecting_2022}

The paper makes three contributions. First, we derive the length exponent of
each layer from the BHH law for minimal networks, and show that the gap
between the exponents of two layers gives the exponent of the density of the
nodes they share (Sec.~\ref{sec:methods}). Second, we measure the length
exponents of the two layers and the density exponent of the substations
independently on the French grid, by Poisson maximum likelihood on every
populated pixel of the population raster, each with a confidence interval from
a spatial block bootstrap (Secs.~\ref{sec:data} ). Third, we compare the
measured node exponent with the predicted one , and we examine the use of the
BHH law in pixels that hold few nodes (Sec.~\ref{sec:results}).

\section{Theory}
\label{sec:methods}

\subsection{Minimal networks}
\label{sec:bhhlaw}

The Steiner minimal spanning tree on a set of points is the shortest network
that connects all of them, allowing auxiliary junction nodes (``Steiner
points'') to be inserted anywhere so as to further reduce the total length.
Its total length obeys the Beardwood--Halton--Hammersley (BHH)
law~\cite{BHH1959,Yukich1998}. For $n$ points drawn independently from a
probability density $f$ of compact support in $d$ dimensions,
\begin{equation}
\frac{L_{n}}{n^{(d-1)/d}}\;\xrightarrow[n\to\infty]{}\;
\beta_{d}\int_{\mathbb{R}^{d}} f(\bm{x})^{(d-1)/d}\,\mathrm{d}\bm{x}
\qquad\text{(a.s.)},
\label{eq:bhh}
\end{equation}
where $L_{n}$ is the network length, $f$ is the \emph{probability} density of
the points ($\int f\,\mathrm{d}\bm{x}=1$), and $\beta_{d}$ is a constant
depending only on the dimension. For $n$ points spread uniformly over a region
of area $A$ in $d=2$, one has $\int f^{1/2}\,\mathrm{d}\bm{x}=A^{-1/2}\cdot
A=\sqrt{A}$, and Eq.~\eqref{eq:bhh} reduces to
\begin{equation}
L\;\sim\;\beta\,\sqrt{n\,A}.
\label{eq:bhh2d}
\end{equation}
We refer to Eq.~\eqref{eq:bhh2d} as the elementary length--count--area
relation from which all the exponents below are built.

Equation~\eqref{eq:bhh2d} is an asymptotic statement, whereas the cells of the
data hold few nodes. In Sec.~\ref{sec:data}, a populated pixel holds one MV/LV
substation in the median.

We use it Eq.~\eqref{eq:bhh2d}, for each grid cell, as a length per unit area
estimator of a network that extends far beyond the cell, so that the sum over
the cells gives the total length. The number of points that must be large is
then that of the network, not that of the cell. This use is tested on
synthetic points in Sec.~\ref{sec:cuttree}.

\subsection{The settled area}
\label{sec:settled}

We write $\rho$ for the number of users per unit area that the network serves
in a cell of area $A$. A cell is essentially never filled uniformly, since its
users occupy only a fraction of it (for a human population, a hamlet, a town
center, or a ribbon of housing along a road), and that occupied area grows as
the cell holds more users. What enters Eq.~\eqref{eq:bhh2d} is therefore not
the area of the cell but the \emph{effective} area $A_{\mathrm{eff}}$ actually
spanned by the served points. We model its growth with $\rho$ by a power law,
\begin{equation}
A_{\mathrm{eff}}(\rho)=A_{0}\,(\rho/\rho_{0})^{\gamma},
\label{eq:Aeff}
\end{equation}
where $A_{0}$ is the effective area at a reference density $\rho_{0}$ and
$\gamma$ is the \emph{settlement exponent}. The different regimes of $\gamma$
have a direct physical reading: $\gamma=0$ is pure densification, where the
footprint stays fixed and only the density of users inside it rises,
$\gamma>0$ is spatial spread, where added users also enlarge the occupied area
(for a human population, suburban sprawl).

\subsection{Density of the shared nodes}
\label{sec:nodes}

The multilevel network we consider is made of two connected layers. The
\textit{upper} layer does not reach the users directly. It reaches the $D$
nodes per unit area that it shares with the \textit{lower} layer, and that in
turn feed the lower layer, as transformer substations do in a power grid. What
sets $D$ is a trade-off between the cost of the nodes and the cost of reaching
them, and the two do not depend on $\rho$ in the same way, so the density that
minimizes their sum is itself a nontrivial power of $\rho$.

Each of the $D$ nodes per unit area serves a share $S=\rho/D$ of the users,
and we let its cost grow as $k\,S^{\theta}$, with $k$ a scale and
$\theta\in[0,1]$ the degree of economies of scale in the nodes: $\theta=0$
gives a fixed cost per node regardless of its load, $\theta=1$ gives a cost
proportional to the load, with no saving from mutualizing it, and $0<\theta<1$
is the intermediate case in which doubling the load less than doubles the
cost. Summed over the $D$ nodes per unit area, the cost of the nodes is
\begin{equation}
C_{\mathrm{nodes}}=D\,k\Big(\frac{\rho}{D}\Big)^{\theta}=k\,\rho^{\theta}D^{1-\theta}.
\label{eq:cnodes}
\end{equation}
Reaching these nodes costs, per unit area,
$C_{\mathrm{links}}=c\,\rho\,D^{-1/2}$, since a node serving an area $1/D$
sits at a typical distance $D^{-1/2}$ from the users it feeds. The two costs
move in opposite directions with $D$, and minimization of their sum,
$C(D)=C_{\mathrm{nodes}}+C_{\mathrm{links}}$, follows from $\mathrm dC/\mathrm
dD=0$,
\begin{equation}
D(\rho)\propto\rho^{\nu}\text{ with } \nu=\frac{1-\theta}{3/2-\theta}.
\label{eq:nu}
\end{equation}
The fixed-cost case $\theta=0$ recovers $\nu=2/3$, consistent with classical
facility-location arguments~\cite{GuseinZade1982,GastnerNewmanPRE}. Positive
$\theta$ lowers this exponent. At $\theta=1$, however, the facility cost is
independent of $D$ and the access cost decreases with $D$: There is no finite
interior optimum without another constraint. Capacity limits, discrete
facilities, nonuniform demand, and shared routes can modify this reasoning.

\subsection{Scaling law of two connected layers}
\label{sec:layer}

\begin{figure}[tbp]
  \centering
  \resizebox{\columnwidth}{!}{\input{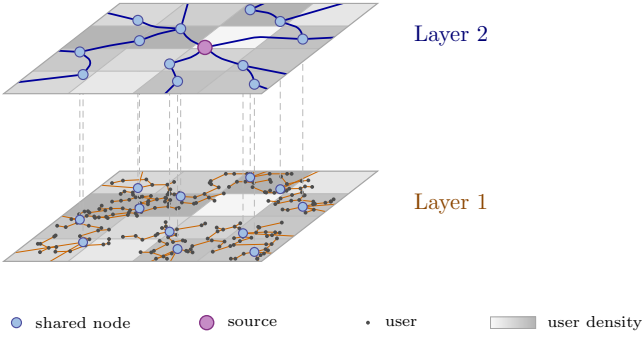}}
  \caption{A two-layer distribution network over the same field of users.
  Layer~1 (orange) reaches the users, layer~2 (blue) reaches the nodes it shares
  with layer~1.}
  \label{fig:multilayer}
\end{figure}

We place ourselves in the case of a network built of two connected layers and
number them from the one that reaches the users, layer~1, to the one that
reaches the shared nodes, layer~2 (Fig.~\ref{fig:multilayer}). To obtain the
length of each layer, we feed Eq.~\eqref{eq:bhh2d},
$\ell\simeq\beta\sqrt{n\,A_{\mathrm{eff}}}$, with the layer's own node count
$n$ and the shared settled area of Eq.~\eqref{eq:Aeff}. The two layers differ
only in $n$: layer~1 reaches every user, so $n_{1}\propto\rho$, whereas
layer~2 reaches only the shared nodes, so $n_{2}\propto
D(\rho)\propto\rho^{\,\nu}$ from Eq.~\eqref{eq:nu}. Inserting these and
collecting the powers of $\rho$ gives
\begin{align}
\ell_{1}(\rho)&\simeq\beta_{1}\sqrt{\rho\,A_{\mathrm{eff}}(\rho)}
\;\propto\;\rho^{(1+\gamma)/2},
\label{eq:bLV}\\[2pt]
\ell_{2}(\rho)&\simeq\beta_{2}\sqrt{D(\rho)\,A_{\mathrm{eff}}(\rho)}
\nonumber\\
&\propto\;\rho^{(\nu+\gamma)/2},
\label{eq:bMV}
\end{align}
where $\ell_{1},\ell_{2}$ are the lengths per unit area of the two layers. The
two-layer exponents are therefore
\begin{equation}
b_{1}=\frac{1+\gamma}{2},
\qquad
b_{2}=\frac{\nu+\gamma}{2}.
\label{eq:exponents}
\end{equation}
Two consequences follow at once. First, the exponent of layer~1 measures the
settlement geometry alone, $\gamma=2\,b_{1}-1$. Second, the gap between the
two layer exponents isolate the multilevel structure,
\begin{equation}
b_{1}-b_{2}=\frac{1-\nu}{2},
\label{eq:gap}
\end{equation}
independently of $\gamma$. Equation~\eqref{eq:gap} is what makes the model
testable. The two length exponents fix $\gamma$ and $\nu$, and solved for
$\nu$ the gap, $\nu=1-2(b_{1}-b_{2})$, predicts the exponent of a third
observable, the density of the shared nodes, with no adjusted parameter. The
settled area cancels in this prediction, so it survives any error on
$A_{\mathrm{eff}}$ that affects the two layers in the same way.

\section{Data and estimation}
\label{sec:data}

\subsection{Data}
\label{sec:datasets}

We now want to compare the theory of Sec.~\ref{sec:layer} with the observed
electricity distribution grid in mainland France. The users are the
inhabitants counted by a gridded population raster, and the cells are the
pixels of that raster. Layer~1 is the LV network which connects the dwellings,
and layer~2 is the MV network which feeds it. The nodes the two layers share
are the MV/LV substations. We write $b_{\mathrm{LV}}$ and $b_{\mathrm{MV}}$
for $b_{1}$ and $b_{2}$.

\begin{figure}[tbp]
  \centering
  \includegraphics[width=\columnwidth]{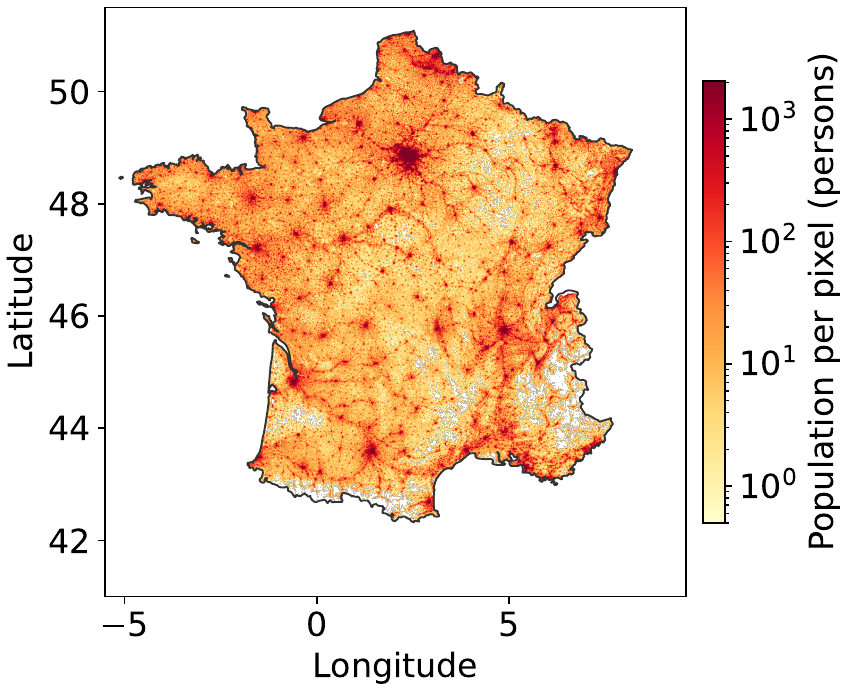}
  \caption{SSP1 2020 population over metropolitan France (excluding Corsica), the
    calibration domain. Logarithmic color scale.}
  \label{fig:francemap}
\end{figure}

\paragraph{Population rasters.}
We use the spatially explicit population projections of
Wang~\textit{et~al.}~\cite{wang_projecting_2022}, which provide gridded
population counts at $30$ arcseconds resolution for each of the five shared
socioeconomic pathway (SSP) scenarios~\cite{oneill_new_2014} at five-year
intervals from 2020 to 2100; we use the 2020 raster of SSP1. At the latitudes
of France a pixel covers $0.54$ to $0.64\ \mathrm{km}^{2}$ (mean $0.59$), and
all counts are converted to densities with the exact area of each pixel. No
gridded population is a ground truth. The SSP rasters distribute census counts
with a random forest model on land cover and settlement
covariates~\cite{Stevens2015}, and they thus carry some information on the
built environment into $\rho$. Figure~\ref{fig:francemap} shows the SSP1 2020
population raster clipped to metropolitan France, excluding Corsica. The
spatial heterogeneity is striking: densely populated urban pixels, as in the
Paris basin and along the coastlines, coexist with large swaths of near-zero
rural population.

\paragraph{Cable network data.}
The LV and MV lines are the open-data export of Agence ORE, the common portal
of the French distribution system operators (DSOs), covering ENEDIS and about
twenty local operators~\cite{ore_lines_opendata}. Corsica is excluded since
its network is not operated by these DSOs. The $19$ operators that publish
their lines serve $99.0\%$ of the delivery
points~\cite{ore_consumption_opendata}, and we restrict the analysis to their
communes (the \emph{main mask}, $98.8\%$ of the population and $858\,078$
populated pixels). The LV files describe the lines that run along the streets,
not the individual service connections ($4.66$ million dead ends for $41.5$
million delivery points). Consequently, the measure of the observed LV
geometry is not the delivery-point count. Its proportionality to population is
an assumption.

\paragraph{Substations.}
The nodes shared between the medium- and low voltage layers are the $964\,940$
MV/LV substations of the same export ($927\,741$ of them
ENEDIS)~\cite{ore_substations_opendata}, each hosting one or more transformers
that step medium voltage down to low voltage. The nodes that feed the MV layer
from above are the $3\,098$ source substations~\cite{ore_sources_opendata}.
The node exponent $\nu$ is measured on the MV/LV substations only. The MV
layer also reaches customers connected directly at medium voltage, which are
not MV/LV substations and are not identified in the open data. They belong to
the enterprise category of the annual electricity consumption per IRIS
district for 2022~\cite{ore_consumption_opendata}, which holds $569\,482$ of
the $41.5$ million delivery points ($1.4\%$) and also contains customers
connected at low voltage. The lines and substations date from 2022, and the
population raster dates from 2020.

\paragraph{Country boundary.}
We use the Natural Earth country boundaries shapefile~\cite{naturalearth} to
delineate metropolitan France and clip the population raster to it.

\paragraph{Cable length per cell.}
For each cable type, we compute the total cable length falling within each SSP
raster pixel as follows:
\begin{enumerate}
  \item Each cable geometry (LineString) is decomposed into its elementary
        straight-line segments.
  \item The great-circle (haversine) length of each segment is
        computed.
  \item Each segment is assigned to the pixel containing its midpoint. Segments
        spanning more than one pixel are subdivided before assignment.
  \item The total cable length in pixel~$i$ is the sum of all assigned segment
        lengths.
\end{enumerate}

This midpoint procedure approximates the allocation of length to cells. It
does not intersect each line exactly with each cell boundary. Assigning whole
segments without any subdivision moves the difference between empirical and
predicted $\nu$ by $0.002$, and removing the pixels within four pixels of the
edge of the study area moves it by less than $0.005$.

\subsection{Maximum likelihood estimation}
\label{sec:estimation}

For pixel $i$, let $A_i$ be its area, $\rho_i$ its population density, and
$y_i$ either line length or substation count. We fit the mean model
\begin{equation}
\mu_i=\mathbb E[y_i\mid\rho_i,A_i] = A_i\,a\,\rho_i^b.
\label{eq:ell_rho}
\end{equation}
All populated pixels enter the fit, including those with no line or no
substation. The $4.4\%$ of masked pixels without population are excluded
because $\log\rho_i$ is undefined there. They contain less than $1\%$ of the
reported line lengths and substation counts. The fit therefore describes the
populated-pixel sample, not every pixel of the territory.

The number of substations in a pixel is a count. At the same resolution, we
modeled it in previous work as a Poisson count of its mean, since its variance
is close to the mean over most of the density range and the share of empty
cells supports this choice~\cite{Emery2026}. The probability of the observed
counts is the product of the Poisson probabilities
$\mu_{i}^{y_{i}}e^{-\mu_{i}}/y_{i}!$. Its logarithm is, up to the terms $-\log
y_{i}!$ that do not depend on $a$ and $b$,
\begin{equation}
\mathcal{L}(a,b)=\sum_{i}\big[y_{i}\log\mu_{i}-\mu_{i}\big],
\label{eq:loglik}
\end{equation}
and the fitted exponent $b$ is the one that maximizes it.

Setting to zero the derivatives of $\mathcal{L}$ with respect to $\log a$ and
to $b$ gives the two likelihood equations
\begin{equation}
\sum_{i}(y_{i}-\mu_{i})=0,
\qquad
\sum_{i}(y_{i}-\mu_{i})\log\rho_{i}=0,
\label{eq:score}
\end{equation}
where $y_{i}-\mu_{i}$ is the residual of pixel $i$, the difference between its
observed and its expected value. The first equation reproduces the observed
total in the fitting sample by construction. The second is a sample
orthogonality condition between the residuals and $\log\rho_{i}$. It does not
show that the residuals are independent of density.

The criterion $\mathcal{L}$ (Eq.~\eqref{eq:loglik}) is concave in $(\log
a,b)$. Its Hessian yields $-\sum_i\mu_i(1,\log\rho_i)^{\top}(1,\log\rho_i)$,
and is negative definite as soon as the pixels do not all share the same
density. The maximizer is then unique when it exists, and every solution of
Eq.~\eqref{eq:score} is this maximizer. It can fail to exist only if no pixel
is positive, or if all positive pixels share a single density. Neither occurs
in the fitting sample. The positive pixels have more than $1.4\times10^{5}$
distinct densities, spanning $1.6$ to $24\,000$ inhabitants per km$^{2}$.

A length is not a count, and the Poisson law is not defined for a continuous
variable. Equations~\eqref{eq:loglik} and~\eqref{eq:score} nevertheless apply
to lengths, because they involve the distribution of $y_{i}$ only through its
mean. A simple picture links the lengths to counts. Let us cut every line into
pieces of a small length $\delta$ and count the pieces of each pixel,
$n_{i}=y_{i}/\delta$. If $n_{i}$ were a Poisson count of mean
$\mu_{i}/\delta$, its log-likelihood would be $\mathcal{L}(a,b)/\delta$, up to
terms that do not depend on $a$ and $b$. It would have the same maximum as
Eq.~\eqref{eq:loglik} for every $\delta$, and its variance would be
$\delta\,\mu_{i}$, proportional to its mean. Neither the size of the pieces
nor the unit of length changes the fitted exponent. Multiplying all lengths by
a positive constant multiplies $a$ by that constant and leaves $b$ unchanged.

Neighboring pixels are not independent, since a settlement or a feeder spans
many of them. This leaves the estimating equations unchanged, but it makes the
uncertainty of the coefficients larger than for independent pixels. We
therefore estimate it with a spatial block bootstrap~\cite{Lahiri2003}. The
territory is cut into squares of $50\,$km. A new sample is built by drawing as
many squares as the territory contains, with replacement, and the three
quantities are refitted on this sample. Repeating the draw $2000$ times gives
$2000$ values of each coefficient, and Table~\ref{tab:exponents} reports the
interval that contains the central $95\%$ of them. The same is done with
squares of $100\,$km. Because the three quantities are refitted on the same
draw, the procedure also gives an interval for any combination of the
coefficients. These intervals describe how much the coefficients vary from one
draw to another.

For any nonnegative pixel quantity $X_j$, define $p_j(\rho)=\mathbb
P(X_j>0\mid\rho)$, the probability that a pixel of density $\rho$ holds this
quantity, and $m_j(\rho)=\mathbb E[X_j\mid\rho,X_j>0]$, its mean over such
pixels. Then $\mathbb E[X_j\mid\rho]=p_jm_j$. We call local exponent of a
function of $\rho$ its slope in logarithmic coordinates, which is the exponent
itself when the function is a power law. Wherever $p_j$ and $m_j$ are positive
and differentiable, the local exponent of the mean is

\begin{equation}
\begin{split}
B_j(\rho)
&=\frac{\mathrm d\log\mathbb E[X_j\mid\rho]}{\mathrm d\log\rho}\\
&=\underbrace{\frac{\mathrm d\log p_j}{\mathrm d\log\rho}}_{h_j(\rho)}
+\underbrace{\frac{\mathrm d\log m_j}{\mathrm d\log\rho}}_{c_j(\rho)}.
\end{split}
\label{eq:decomp}
\end{equation}
For line length, $X_j$ is the length per unit area. The same decomposition
applies to the number of substations per unit area. Presence means that the
cell contains an observed line or node, not necessarily that its inhabitants
are served by that object. The positive-cell term $c_j$ can include routing,
topology, and boundary effects. We keep the empty cells and report the
exponent of the mean across all cells throughout, since it gives the expected
length of a cell from its population density alone.

\section{Results}
\label{sec:results}

\subsection{Exponents at the pixel scale}
\label{sec:pixel}

\begin{figure*}[tbp]
  \centering
  \includegraphics[width=\textwidth]{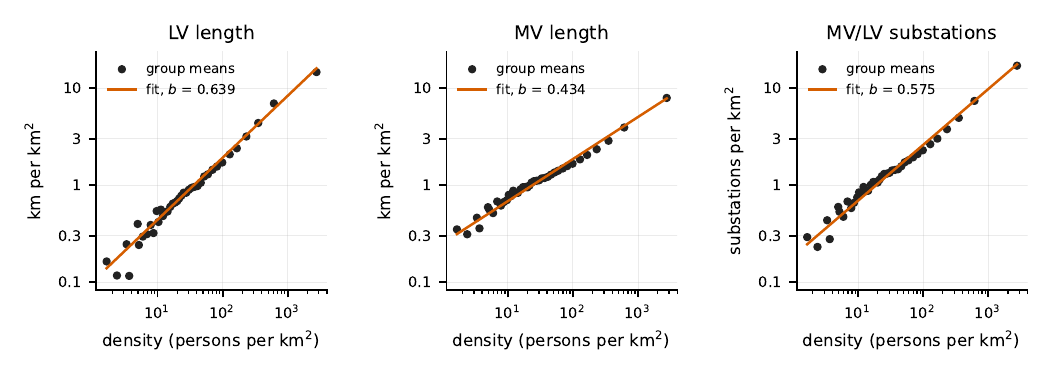}
  \caption{Length per unit area of the LV and MV layers and density of
  MV/LV substations against population density, over the main mask, empty
  populated pixels included. Points: means of $40$ equal-count groups of the populated
  pixels, shown for reading only. Lines: the power law~\eqref{eq:ell_rho}
  fitted by maximum likelihood on all the populated pixels, with its
  exponent $b$.}
  \label{fig:fits}
\end{figure*}

\begin{table}[tbp]
\caption{Global effective exponents at the pixel scale, fitted by
Poisson maximum likelihood on populated pixels. Brackets are $95\%$ spatial block
bootstrap intervals for $50$ and $100\,$km squares ($2000$ replicates). The quantity
$\gamma_{\mathrm{emp}}$ is an apparent geometric
parameter under the reference assumptions.
The empirical estimates $\nu_{\mathrm{emp}}$
uses the two fitted length coefficients. The last row is the
difference $\Delta$, with
its interval obtained from the same replicates.}
\label{tab:exponents}
\footnotesize
\setlength{\tabcolsep}{0.5pt}
\begin{ruledtabular}
\begin{tabular}{lccc}
Quantity & Estimate & $50\,$km & $100\,$km \\
\hline
$b_{\mathrm{LV}}$ & $0.639$ & $[0.621, 0.659]$ & $[0.613, 0.672]$ \\
$b_{\mathrm{MV}}$ & $0.434$ & $[0.416, 0.452]$ & $[0.409, 0.459]$ \\
$\nu$ & $0.575$ & $[0.563, 0.588]$ & $[0.556, 0.598]$ \\
$\gamma_{\mathrm{emp}}=2b_{\mathrm{LV}}-1$ & $0.277$ & $[0.243, 0.318]$ & $[0.226, 0.345]$ \\
$\nu_{\mathrm{emp}}=1-2(b_{\mathrm{LV}}-b_{\mathrm{MV}})$ & $0.590$ & $[0.524, 0.651]$ & $[0.502, 0.668]$ \\
$\Delta = \nu - \nu_{\mathrm{emp}}$ & $-0.015$ & $[-0.076, 0.049]$ & $[-0.096, 0.079]$ \\
\end{tabular}
\end{ruledtabular}
\end{table}

Figure~\ref{fig:fits} shows the two line-length relations and the
substation-density relation, and Table~\ref{tab:exponents} reports their
global effective exponents. A single power law captures the trend of each
quantity over the range of the group means, from about $2$ to $3\,000$ persons
per $\mathrm{km}^{2}$, with $b_{\mathrm{LV}}=0.639$, $b_{\mathrm{MV}}=0.434$
and $\nu=0.575$.

\begin{figure}[tbp]
  \centering
  \includegraphics[width=\columnwidth]{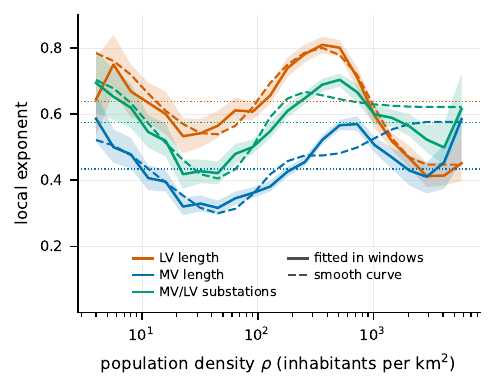}
  \caption{Local exponent of the LV length, the MV length, and the substation density as functions of the population density. Full lines:
  power law fitted by Poisson maximum likelihood inside windows
  $[c/2,2c)$ centred on $c$; shaded bands: $95\%$ intervals from the block
  bootstrap on $50\,$km squares. Dashed lines: derivative with respect to
  $\log\rho$ of a smooth curve with seven free parameters, fitted to the
  logarithm of the mean on all pixels.
  Dotted lines: the constant exponents of Table~\ref{tab:exponents}.}
  \label{fig:local_slope}
\end{figure}

The local exponent is not constant. Fitting the power law inside windows of
density $[c/2,2c)$ gives the full lines of Fig.~\ref{fig:local_slope}: the
exponents range from $0.30$ to $0.80$ and follow a wave common to the three
quantities below $1000$ inhabitants per $\mathrm{km}^2$. A smooth curve fitted
on all pixels gives the same shape (dashed lines).

A single exponent is therefore a two-parameter summary, not a local law. It
nevertheless predicts total lengths out of sample. We hold out whole $100\,$km
squares in five random folds, fit the model on the other four folds, and
predict the held-out squares. Summed over all squares, the predictions are
within $1\%$ of the observed totals for the LV length, the MV length, and the
number of substations. The error on the total of a single square is larger:
its median absolute value is $18\%$, $14\%$, and $17\%$, and the
root-mean-square error of its logarithm is $0.27$, $0.19$, and $0.24$. A
smooth curve in $\log\rho$ lowers the latter by less than $0.01$. A fixed
length per inhabitant, fitted in the same folds, gives $0.56$, $0.70$, and
$0.61$.

The exponents are sublinear ($b<1$), confirming that cable infrastructure
scales more slowly than population, consistent with urban scaling
theory~\cite{Bettencourt2007}. On the group means, the presence part of
Eq.~\eqref{eq:decomp} accounts for $35\%$ of $b_{\mathrm{LV}}$, $51\%$ of
$b_{\mathrm{MV}}$ and $56\%$ of $\nu$ at the pixel scale. Much of it is the
arithmetic of counting discrete objects in small cells. These shares fall to
$31\%$, $45\%$, and $46\%$ on a $1\,$km grid, where fewer cells are empty.
These percentages describe those group-mean diagnostics and should not be
treated as an exact decomposition of the fitted coefficients. They
nevertheless show that network presence is a major component of the observed
density dependence.

\subsection{Global comparison with prediction}
\label{sec:gamma_nu}

Through Eq.~\eqref{eq:exponents}, the LV exponent gives the settlement
exponent, $\gamma=2b_{\mathrm{LV}}-1=0.277$ (Table~\ref{tab:exponents}). It is
positive and well below one. It indicates mild spatial spread, since within a
cell, the settled area grows sublinearly with population, and the density of
users in this area rises. The area of Eq.~\eqref{eq:Aeff} measures how the
users occupy the pixel, not how much of it is built, it is thus hard to
measure empirically.

A positive $\gamma$ means that the settled area of a pixel grows with its
population, which cannot go on at every density since that area is bounded by
the pixel. We stress that Eq.~\eqref{eq:Aeff} is a local power-law
approximation, valid over the range of pixel populations spanned by the
calibration data, rather than a relation holding uniformly for all $\rho$.
Since the inhabited area cannot exceed the pixel area,
$A_{\mathrm{eff}}(\rho)\le A_{\mathrm{pix}}$, the power law \eqref{eq:Aeff}
must eventually saturate as $\rho\to\infty$, and capturing this crossover
would require promoting $\gamma$ to a density-dependent running exponent
$\gamma(\rho)\to 0$ in the high-density limit. The sublinear growth of the
settled area with population is not \emph{ad hoc}: sublinear area--population
relations are well documented at the scale of whole
cities~\cite{Bettencourt2013,MolineroThurner2021}.

The model is over-determined and therefore falsifiable. Through
Eq.~\eqref{eq:gap}, the node exponent fixed by the two length exponents
predicts the scaling of a \emph{third}, independent observable, the number of
MV/LV substations per cell,
$\nu_{\mathrm{pred}}=1-2\,(b_{\mathrm{LV}}-b_{\mathrm{MV}})=0.590$, against
$\nu=0.575$ counted on the substations. The node exponent obtained by
\emph{counting substations} therefore agrees within the resolution of the test
with the one inferred from the \emph{gap between the two length exponents},
and the loop closes across three quantities recorded separately, the LV
length, the MV length, and the substation positions. In other words, the LV
length of a pixel exceeds its MV length by a factor that grows as the square
root of the number of users per substation, $\rho/D$, as Eqs.~\eqref{eq:bLV}
and~\eqref{eq:bMV} gives for two trees over the same area. The prediction
needs no adjusted parameter.

Urban MV feeders are built as loops that give each substation a backup path,
and the cables of several feeders share the same streets, so the minimal tree
picture does not describe the city centers. The agreement over the whole
territory is thus a property of exponents averaged over the density range.

\subsection{Minimal networks in cells that hold few nodes}
\label{sec:cuttree}

\begin{figure*}[tbp]
  \centering
  \includegraphics[width=\textwidth]{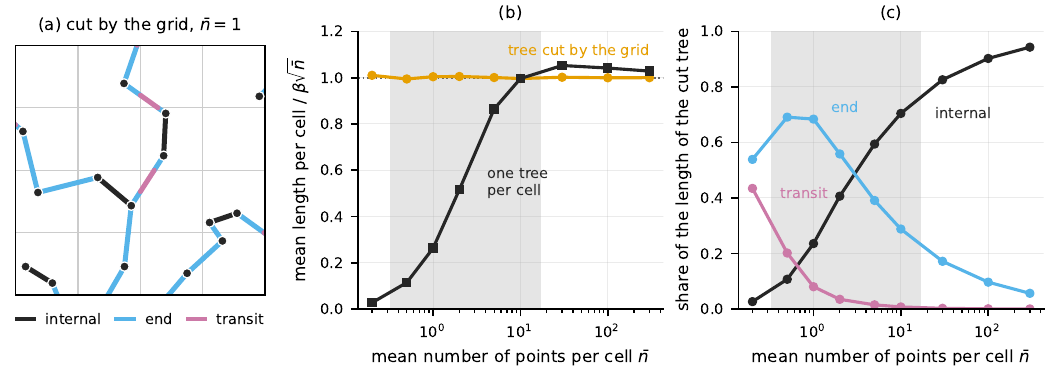}
  \caption{Length read in a cell for uniform random points on a grid of
  cells of unit area, with a mean of $\bar n$ points per cell. (a) Minimum
  spanning tree built on all the points and then cut by the grid, for
  $\bar n=1$. The length in a cell is internal when the edge has its two
  ends in the cell, end when it has one, and transit when it has none.
  (b) Mean length per cell divided by $\beta\sqrt{\bar n}$, with
  $\beta=0.647$, for this tree and for trees built on the points of each
  cell alone. (c) Shares of the three kinds of length in the tree cut by
  the grid. The gray bands cover the mean numbers of MV/LV substations per
  pixel of the density classes of Sec.~\ref{sec:data}, from $0.3$ to
  $17$.}
  \label{fig:cuttree}
\end{figure*}

The exponents above rest on Eq.~\eqref{eq:bhh2d}, used as a length per unit
area in pixels that hold few nodes (Sec.~\ref{sec:bhhlaw}).
Figure~\ref{fig:cuttree} tests this use with uniform random points on a grid
of cells of unit area, where $\bar n$ is the mean number of points per cell.
The test is made with the minimum spanning tree, which obeys
Eq.~\eqref{eq:bhh} with its own constant~\cite{Yukich1998}. The tree is built
on all the points and then cut by the grid to illustrate our method
[Fig.~\ref{fig:cuttree}(a)]. Its mean length per cell stays within $1\%$ of
$\beta\sqrt{\bar n}$ from $\bar n=0.5$ to $\bar n=300$
[Fig.~\ref{fig:cuttree}(b)]. The tree connects all the points, so it is in the
asymptotic regime whatever the size of the cells, and for points of uniform
density its length per unit area is the same everywhere. At $\bar n=1$, $76\%$
of this length belongs to edges that leave the cell or only cross it
[Fig.~\ref{fig:cuttree}(c)], and a tree built on the points of the cell alone
has $26\%$ of $\beta\sqrt{\bar n}$. The asymptotic regime is not reached when
we build the minimal network only within the cell. Two conditions follow. The
count that enters Eq.~\eqref{eq:bhh2d} is the mean number of nodes of a cell
at a given density, and the length of a cell includes the lines that leave it
or cross it. In other words, given a density and the area of a cell, we take
Eq.~\eqref{eq:bhh2d} as the fraction of an ideal global network that is
located within the area of the cell. Both are met in
Sec.~\ref{sec:estimation}, where the empty pixels are kept and the length of a
pixel is that of all the lines found in it.

\section{Discussion}
\label{sec:discussion}

The model reaches its first aim. Two exponents and two amplitudes, fitted on a
population raster, give the length of the LV and MV layers of a pixel from its
population density alone, and they reproduce the total lengths of the French
grid by construction. The exponents are consistent with the minimal picture of
Sec.~\ref{sec:layer} in two respects. The LV exponent implies a settled area
that grows sublinearly with the density, and the gap between the two layers
predicts the density exponent of the substations within the resolution of the
test.

The power-law form of Eq.~\eqref{eq:ell_rho} is a two-parameter summary, not a
locally exact law: the exponent fitted within density windows ranges from
$0.30$ to $0.80$ (Fig.~\ref{fig:local_slope}). It nonetheless predicts the
total length of a $100\,$km square to within $0.8\%$ (LV), $0.2\%$ (MV), and
$0.4\%$ (substations), while the error on the total of a single square is
larger, with a root-mean-square logarithmic value of $0.27$, $0.19$, and
$0.24$ respectively. The unconditional exponent $b$ itself adds a presence
part and a geometric part [Eq.~\eqref{eq:decomp}]. We keep the unconditional
exponent because it is the one that predicts a length map from a population
map.

Finally, the node exponent lies below the $2/3$ of facilities sited by travel
distance~\cite{GastnerNewmanPRE,um_scaling_2009}. The benchmark of
Sec.~\ref{sec:nodes} attributes this to economies of scale in the substations.

The present study concerns one country, one pair of voltage levels, 2022
network geometries, and a 2020 population field. Direct MV customers are
absent from the shared-node count, and LV service connections are absent from
the mapped line geometry. Replacing the $30$-arcsecond pixels by a $1\,$km
Lambert-93 grid changes the three fitted coefficients by less than $0.005$ and
$\Delta$ by $-0.011$ $[-0.014,-0.008]$. Merging cells lowers the three
coefficients. From $1$ to $32\,$km, $b_{\mathrm{LV}}$ goes from $0.634$ to
$0.544$, $b_{\mathrm{MV}}$ from $0.433$ to $0.364$, $\nu$ from $0.573$ to
$0.495$, and $\Delta$ from $-0.026$ $[-0.093,0.038]$ to $-0.147$
$[-0.226,-0.054]$. The relation is therefore supported at the kilometer scale
only. Spatial bootstrap intervals quantify resampling uncertainty, not the
errors of the population model, the lines missing from the geometry, or the
assumptions of the geometric reference.

The test of Fig.~\ref{fig:cuttree} is limited to uniform points. In synthetic
fields whose density varies from cell to cell, the length of the tree is
$98\%$ to $100\%$ of the sum of Eq.~\eqref{eq:bhh2d} over the cells, but the
exponent fitted on the cells exceeds $1/2$ by $0.01$ to $0.09$. Points grouped
in clusters inside the cells lower this exponent to $0.28$ and the length to
$53\%$ of the sum, a situation that the effective area of Eq.~\eqref{eq:Aeff}
is meant to describe. The arrangement of the nodes within and between pixels
can therefore shift the exponents.

\section{Conclusion}
\label{sec:conclusion}

We derived the length exponent of each layer of a multilevel distribution
network from the BHH law, in a minimal form in which the layers share the
settled area and differ only in the number of nodes they reach. We tested it
on the complete low and medium voltage grid of metropolitan France, at the
scale of the population raster pixel, with the population density as the only
input. The LV exponent implies a settled area that grows sublinearly with the
density, and the two length exponents predict the density exponent of the
substations, $0.590$ against $0.575$ measured, within a resolution of about
$0.1$. The difference from the observed exponent, $-0.015$, has a $95\%$
interval of $[-0.076,0.049]$. This agreement holds for the whole territory,
and coexists with substantial presence effects, curvature, and reported
failures within density windows. All exponents depend on the resolution,
through a dependence of the length of a pixel on the density around it.

The distribution network is associated with concrete and economically pressing
issues, since estimating its current extent and future evolution under
different demographic scenarios is essential for quantifying the material,
investment, and maintenance requirements of the energy transition. The
calibrated $\ell(\rho)$ relation, applied pixel by pixel to the SSP population
projections~\cite{wang_projecting_2022,oneill_new_2014}, turns a demographic
scenario into a spatially resolved estimate of distribution-network length,
and hence of the associated material and investment
requirements~\cite{Emery2026}. The population field is a constraint of such a
use. The SSP rasters spread the census counts over the territory with
covariates of the built environment~\cite{Stevens2015}, and other population
fields~\cite{ghs_pop_2023,filosofi2019} disagree with them on which sparse
pixels are populated. The exponents are thus attached to the field they were
calibrated on, and the model should be applied with the same field, which for
projections means the SSP rasters themselves.

\paragraph*{\textbf{Data availability}}---
Raw network, substation, population, and ancillary data are available from
their providers~\cite{ore_lines_opendata,ore_substations_opendata,
ore_sources_opendata,ore_consumption_opendata,wang_projecting_2022,BDTOPO}.

\paragraph*{\textbf{Acknowledgments}}---
We would like to thank Joseph Le Bihan for helpful discussions.
\textbf{Funding}--- This work has benefited from a government grant managed by
the Agence Nationale de la Recherche under the France 2030 program, reference
ANR-22-PERE-0003.

\bibliography{biblio}

\end{document}